\documentclass[useAMS,usenatbib]{mn2e}
\usepackage{graphicx}

\title[High-frequency QPOs in GRS 1915+105]
{High-frequency Quasi-Periodic Oscillations from GRS 1915+105 in its C state}
\author[T. Belloni et al.]{T. Belloni$^{1}
$\thanks{E-mail:belloni@merate.mi.astro.it},
P. Soleri$^{1,2,3}$, P. Casella$^{1,3}$, M. M\'endez$^{4}$, S. Migliari$^{5,3}$
\\
$^{1}$ INAF-Osservatorio Astronomico di Brera, Via E. Bianchi 46, I-23807 Merate
(LC), Italy\\
$^{2}$ Universit\`a degli Studi di Milano, Dipartimento di Fisica, Via Celoria 16, 20133, Milano, Italy\\
$^{3}$ Astronomical Institute ``Anton Pannekoek'', University of Amsterdam, and Center for High Energy Astrophysics, Kruislaan 403,\\ 1098 SJ, Amsterdam, The Netherlands\\
$^{4}$ SRON, Netherlands Institute for Space Research, Sorbonnelaan 2, 3584 CA Utrecht, The Netherlands\\
$^{5}$ Center for Astrophysics and Space Sciences, Code 0424, University of California at San Diego, La Jolla, CA 92093, USA\\
}

\begin{document}

\date{
Accepted ... Received ...; in original form ...}

\pagerange{\pageref{firstpage}--\pageref{lastpage}} \pubyear{2005}

\maketitle

\label{firstpage}

\begin{abstract}

We report the results of a systematic timing analysis of RXTE observations of GRS 1915+105 when 
the source was in its variability class $\theta$, characterized by alternating soft and hard states on 
a time scale of a few hundred seconds. The aim was to examine the high-frequency part of the
power spectrum in order to confirm the hecto-Hertz Quasi-Periodic Oscillations (QPO) previously reported from observations
from mixed variability behaviours. During the hard intervals (corresponding to state C in the 
classification of Belloni et al., 2000, A\&A, 35, 271), we find a significant QPO at a 
frequency of $\sim$170 Hz, although
much broader (Q$\sim$2) than previously reported. No other significant peak is observed at 
frequencies $>$30 Hz. A time-resolved spectral analysis of selected observations shows that the
hard intervals from class $\theta$ show a stronger and steeper ($\Gamma$=2.8-3.0) power-law component than hard intervals from other classes.
We discuss these results in the framework of hecto-Hertz QPOs reported from GRS 1915+105 and other black-hole binaries.

\end{abstract}


\section{Introduction}

Since the launch of the NASA satellite Rossi X-Ray Timing
Explorer (RXTE) in 1995, a number of
high-frequency Quasi-Periodic Oscillations (HFQPOs) have been discovered in
several Low-Mass X-ray Binaries (van der Klis 2005). Although many different
theories have been formulated in the last decade (for a review see van der Klis 2005), using
different approaches in order to explain the observed phenomenology, no
consensus has been reached at present on the physical origin of these
features, and work is still in progress in order to improve our knowledge of
their behaviour.

In particular, HFQPOs have been discovered only in a few Black Hole
Candidates (BHCs). They often appear as transient and subtle features in the
Fourier power density spectra (PDS), with centroid frequencies of tens to hundreds
of Hz and low rms amplitude values
($\sim$1-3\%) strongly dependent on the energy range (see e.g. Morgan, Remillard \& Greiner 1997). 
Three BHCs show a single HFQPO: XTE J1650-500 (250 Hz, Homan et al. 2003),
4U 1630-47 (variable centroid frequency, Klein-Wolt, Homan \& van der Klis 2004) 
and XTE J1859+226 (190 Hz, Cui et
al. 2000), while other four sources exhibit pairs of HFQPOs: GRO J1655-40 (300 and 450 Hz, 
Remillard et al. 1999, Strohmayer 2001a), XTE J1550-564
(184 and 276 Hz, Homan et al. 2001, Miller et al. 2001, Remillard et al. 2002a), H1743-322
(165 and 241 Hz, Homan et al. 2005) and GRS 1915+105 (41 and 69 Hz, Morgan et al. 1997, 
Strohmayer 2001b).
In many systems, the observed centroid frequencies appear rather stable, showing only
small variations between different observations. As first noted by Abramowicz \& Kluzniak (2001) for the case of GRO J1655-40,
when two QPOs are observed, they appear at frequencies consistent with a simple ratio (2:3 or 3:5).
As mentioned above, these features in the power spectra are very weak and observed
only in a few cases. The relation between the detection of HFQPOs and source states
is not completely clear, although it appears that most of them are observed during the
Soft Intermediate State (see Belloni et al. 2005, Homan \& Belloni 2005). At any rate, most of the QPO detections listed above are from observations when no type-C low-frequency QPOs are observed (see Casella, Belloni \& Stella 2005 and references therein for a definition of these QPO types).

GRS 1915+105 is a very bright BHC that appeared in the X-ray sky in 1992 (Castro-Tirado et al. 1992) and has been bright in the sky ever since (see Fender \& Belloni 2004 for a review). It is a very peculiar source which shows strong variability on time scales of seconds to months, unique to this system (see Belloni et al. 2000). 
The first HFQPO detected in this system was at a frequency of $\sim$69 Hz and was observed
in very few RXTE observations in 1996 (Morgan et al. 1997). The QPO  has a hard energy spectrum, with an integrated fractional rms that increased from 1.5\% below 5 keV up to 6\% above 13 keV.
The QPO appeared during observations where strong quasi-periodic oscillations with a period of 15-20 seconds were observed. From the light curve and the hardness-intensity diagrams, it can be concluded that during these observations GRS 1915+105 was in its $\gamma$ variability mode (see Belloni et al. 2000; Belloni, M\'endez \& S\'anchez-Fern\'andez 2001). Two of these observations were re-analyzed by Belloni et al. (2001), who found that in one observation (1996 May 5th) the 65 Hz QPO changed properties as a function of the phase of the $\sim$15 s oscillation, while in the other observation (1996 May 14), no significant 65 Hz QPO, but a 27 Hz QPO was visible at certain phases of the $\sim$15 s oscillation. A second QPO peak at a frequency of $\sim$41 Hz was detected by Strohmayer (2001b) by analyzing RXTE data above 13 keV; unlike the 27 Hz one, this QPO was also detected simultaneously with a 67 Hz peak. It is interesting to note that 27:41:67 are in 2:3:5 ratio.

Remillard et al. (2002b) reported the detection of two additional HFQPO peaks from GRS 1915+105, at a frequency of 164 and 328 Hz, with the second one being consistent with the second harmonic of the first.
These peaks were obtained from a selection in X-ray hardness and intensity in a set of RXTE observations from 1997 September. The selected parts of the observations were all from the C state of GRS 1915+105 (see Belloni et al. 2000), corresponding to the Hard Intermediate State (HIMS) in other transients (see Belloni et al. 2005, Homan \& Belloni 2005). In this state, C-type QPOs are always observed and indeed are apparent in the power spectra from Remillard et al. (2002b). 
These detections are the only case of harmonically related HFQPOs and one of the rare cases of simultaneous presence of HFQPOs and type-C QPOs. Assuming they are different from the 27, 41 and 69 Hz peaks (as they appear in different states, this is not at all guaranteed), this source would show at least four independent HFQPOs, three of them appearing in 2:3:5 ratio, and two higher-frequency ones appearing in 1:2 ratio. 

The work presented in this paper is aimed at searching for high-frequency QPOs in 
GRS 1915+105 following a procedure similar to that of Remillard et al (2002b), 
extending the search to a broader
data-set, in order to better constrain their phenomenology.
Many observations in the sample by Remillard et al. (2002b) belong to variability class $\theta$ (see Belloni et al. 2000). This class is characterized by alternating intervals of the soft state A and the harder state C; it is particularly interesting to note that the C-state intervals appear to be peculiar compared to other variability classes, being characterized by a rather high count rate and low values of  X-ray hardness (see Belloni et al. 2000).
We selected all available RXTE observations 
in the time interval 1996 April to 1999 March
from the public archive which could be classified as class $\theta$ and performed a search for HFQPOs similar to that from Remillard et al. (2002b). 

\section{Class $\theta$: colors and spectra}

\begin{figure} \resizebox{\hsize}{!}{\includegraphics{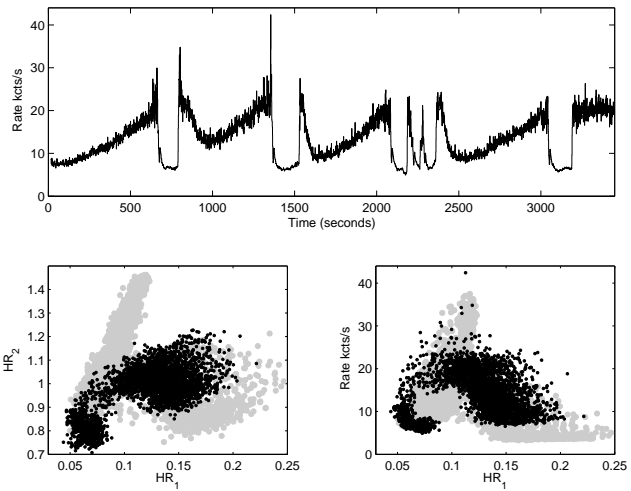}}
\caption{Top panel: PCA 2-13 keV $\theta$ light curve of GRS 1915+105
from 1997 Sep 5. Bottom panels: corresponding Color-Color 
diagram and Hardness-Intensity Diagram (see Belloni et al. 2000). The 
gray points are from an observation of 1997 Sep 9, when the source was
in its $\beta$ class. The time resolution of all plots is one second.
}
\label{lego}
\end{figure}

Variability class $\theta$ as defined in Belloni et al. (2000) includes the 
only observations when the source does not go through state B. A typical
RXTE/PCA light curve (2--13 keV) 
is shown in the top panel of Fig. \ref{lego}: the characteristic
`M' shape is caused by the alternating C-state (high rate and variable) and
A-state (low rate and quiet) intervals. The corresponding 
color-color diagram, defined as in Belloni et al. (2000) is shown in the bottom
left panel of Fig. 1. State C corresponds to the cloud of points centered around
(0.15,1.0), while state A is represented by the soft points in the lower
left of the diagram. As comparison, we plot in light gray also the points
corresponding to a typical observation of class $\beta$. One can see
that, in addition to the lack of the characteristic `finger' going to 
state B, the state-C points of class $\theta$ are displaced with 
respect to class $\beta$. The differences can be clearly seen in the
Hardness-Intensity Diagram (see Belloni et al. 2000) shown in the
bottom right panel of Fig. \ref{lego}. Notice in particular how the $\theta$
points corresponding to state C have a much higher rate than those of
class $\beta$. This class shows distinctive characteristics and has not
been analyzed as deeply as, for instance, class $\beta$ (see e.g. 
Mirabel et al. 1998; Markwardt et al. 1999; Migliari \& Belloni 2003). Notice that the
$\theta$ and $\beta$ observations shown in Fig. \ref{lego} are separated
only by four days, indicating that the source can transit rather easily between
these two `modes'.

\begin{figure} \resizebox{\hsize}{!}{\includegraphics{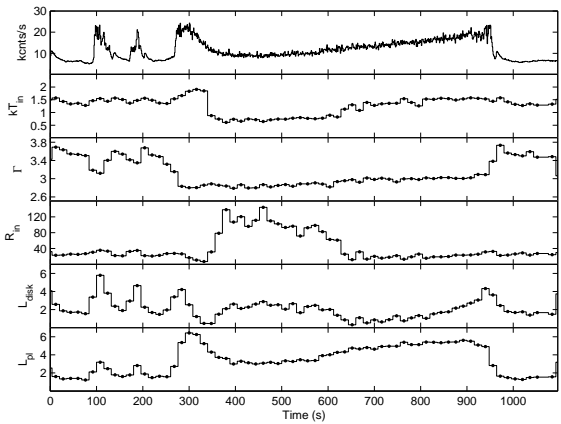}}
\caption{Panels from top to bottom: (1) PCA count rate (kcnts/s, 1s
bin size); (2) inner disc temperature (keV); (3) power-law photon
index; (4) inner disc radius (km); (5) bolometric disc luminosity 
($10^{38}$erg/s); (6) 3-25 keV power-law luminosity ($10^{38}$erg/s)
See Migliari \& Belloni (2003) for more details.
}
\label{simone}
\end{figure}

\begin{figure}\resizebox{\hsize}{!}{\includegraphics{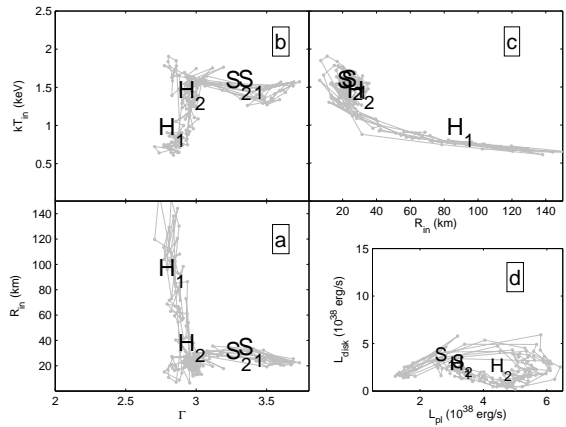}}
\caption{Correlations between the spectral parameters of Fig. 
\ref{simone}. The axes have
the same limits as Fig. 3 in Migliari \& Belloni (2003), with the exception
of the power-law luminosity, which in this case exceeds that limit.
The $H_n$ and $S_n$ labels indicate average values corresponding to the
rate-hardness selection in Sect. 3.1.
}
\label{simone2}
\end{figure}

In order to examine the possible nature of these differences in the
spectral domain, we applied to the PCA data of the
observation of 1997 Sep 5th the same
procedure for time-resolved spectral extraction described in Migliari
\& Belloni (2003). The energy spectra, with a 16s time resolution, were
fitted with a simple model consisting of the superposition of a
disc-blackbody (Mitsuda et al. 1984) and a power law. The interstellar
absorption was fixed to 7$\times 10^{22}$cm$^{-2}$. The results for a full
A-C-A cycle are shown in Figures \ref{simone} and \ref{simone2}
(in the same format as the corresponding figures for $\beta$ observations 
in Migliari \& Belloni 2003). Clear differences with the
$\beta$ case are evident: the disc component is fainter,
and in state C the energy spectrum is dominated by a steeper power-law
component. Notice that this means that the observed count rate in 
the C-state interval is dominated by the power law more than the power-law flux
indicates, given its spectral steepness.

In summary, the major peculiarities of this variability class are the
absence of a state B and the
strength and softness of the power-law component during state C.

\section{Timing/hardness analysis}

We examined all PCA light curves (with 1 second binning) of RXTE observations of GRS 1915+105 
corresponding to PCA Gain Epoch 3 (1996 April 15 to 1999 March 22).
For our analysis, we follow a procedure similar to that adopted by Remillard et al. (2002b). We divide each observation in intervals 16 seconds long. For each interval, we compute a total count rate from all Proportional Counter Units (PCUs) in the PHA channel range 0-79 (corresponding to 2-30 keV) and a hardness ratio defined as the ratio of counts in channel range 36-79 (13-30 keV) to those in channel range 0-35 (2-13 keV). As the source is very strong, we apply only an average background subtraction using fixed values for all the bands: this procedure does not modify significantly the shape of the hardness-intensity diagram. For each interval, we also compute a power spectrum from the 36-79 channel range (13-30 keV) covering the frequencies 0.0625 to 1024 Hz,
which we normalize to squared fractional rms (see Belloni \& Hasinger 1990). As a precise estimate of the noise level due to Poissonian statistics is difficult to make, especially for such a strong source, and as the detection of high-frequency features depends crucially on this estimate, we do not subtracted the Poissonian level, but included it in our model as an additive constant value that is free to vary.

\subsection{Analysis of class $\theta$ observations}

As this variability class is characterized by a very clear pattern, we selected all observations which belonged to this class.
The list of observations is shown in Table \ref{mariano}.

%                                                       One column table
%-----------------------------------------------------------------------
\begin{table}
\begin{tabular}{ll}
\hline
\textbf{Obs. ID} & \textbf{Start Date} \\
\hline
10408-01-15-05 & 1996 Jun 16 14:23\\
10408-01-15-04 & 1996 Jun 16 15:59\\
10408-01-15-00 & 1996 Jun 16 17:35\\
10408-01-15-01 & 1996 Jun 16 19:11\\
10408-01-15-02 & 1996 Jun 16 20:47\\
10408-01-15-03 & 1996 Jun 16 22:23\\
10408-01-16-00 & 1996 Jun 19 14:25\\
10408-01-16-01 & 1996 Jun 19 16:01\\
10408-01-16-02 & 1996 Jun 19 17:37\\
10408-01-16-03 & 1996 Jun 19 19:13\\
10408-01-16-04 (\#2)& 1996 Jun 19 20:49\\ 
20402-01-45-02 & 1997 Sep 05 4:50\\
20186-03-02-00 & 1997 Sep 15 4:21\\
20186-03-02-01 & 1997 Sep 15 18:48\\
20186-03-02-02 & 1997 Sep 16 2:34\\
20186-03-02-03 & 1997 Sep 16 8:18\\
20186-03-02-04 (\# 1, 2, 3) & 1997 Sep 16 18:46\\
20187-02-04-00 (\# 2) & 1997 Oct 05 8:15\\

\hline
\end{tabular}
\caption[]{\footnotesize The two groups of selected $\theta$ observations. \#
  indicates the orbit number.}
\label{mariano}
\end{table}
%-----------------------------------------------------------------------

\begin{figure}
\begin{center}
\includegraphics[angle=0,width=8.5cm]{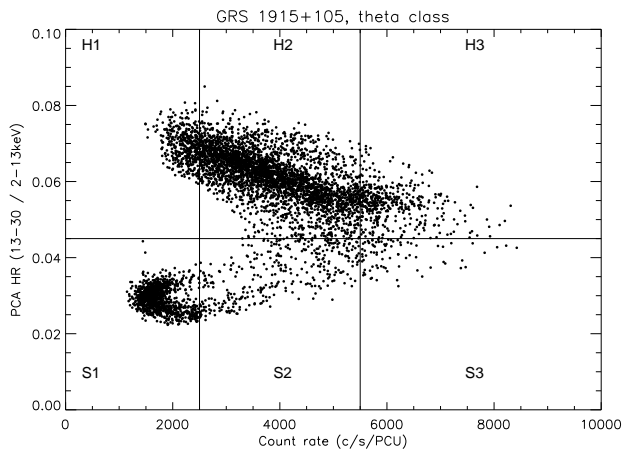}
\end{center}
\caption{Hardness-intensity diagram for GRS 1915+105 using the RXTE/PCA observations
for variability class $\theta$ listed in Table \ref{mariano}.
Each point corresponds to an accumulation time of 16 s. The hardness
  (HR) is obtained as the ratio of source counts in the 13-30 keV band over those in the 2-13 keV band, while the intensity corresponds to 2-30 keV.
  The plot is divided in six regions according to the criteria of Remillard et al. (2002b).}
\label{theta}
\end{figure}

\begin{figure}
\begin{center}
\includegraphics[angle=0,width=8.5cm]{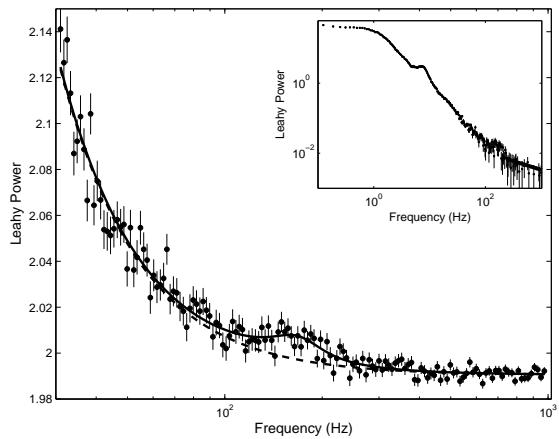}
\end{center}
\caption{Power spectrum for the H2+H3 selection for the $\theta$ class observations (see Fig. 
\ref{theta}). The main panel shows the high-frequency portion. The thick line represents the
best fit model (see text), while the dashed line is the same model with the Lorentzian component
excluded. The inset shows the full power spectrum, after subtraction of the best-fit Poissonian
level.}
\label{pds_theta}
\end{figure}

In Figure \ref{theta}, we show the total hardness-intensity diagram,
as obtained from the complete data-set in Table \ref{mariano}. The total number of points (with 16 s exposure) is 5741.
We use the same time resolution, energy
ranges and normalization to PCU2 adopted by Remillard et al. (2002b). 
Two main regions can be seen in the plot: the region at higher hardness corresponds to state C, while the smaller region at lower count rate and hardness corresponds to state A. As expected, for these observations no state B is observed (see Belloni et al. 2000). 
The fast transitions between states A and C are represented mostly by the points in region S2 (see Fig. \ref{theta}). Following Remillard et al. (2002b), we divided the HID in six sections,
marked in Fig. \ref{theta}. In Fig. \ref{simone2} we marked the position of the average spectral
parameters corresponding to these regions (for S3 and H3, no points were present in this 
particular observation).
We then averaged the power spectra for the two high-rate hard sections (H2 and H3) and for the union of the two sections (H2+H3). The number of points in the H2 and H3 regions are 3279 and 420 respectively, corresponding to an exposure of 52464 and 6720 seconds.
The resulting power spectrum for H2+H3, limited to its high-frequency part ($>$30 Hz),
is shown in Fig.\ref{pds_theta}.

We first fitted the three power spectra with a model consisting of a power law for the continuum noise and a flat component for the Poissonian component. 
We obtained reduced $\chi^2_\nu$ values of 1.40, 1.14
and 1.33 for the regions H2, H3 and H2+H3 respectively.

A visual inspection of the
residuals revealed the presence of a power excess around $\sim$ 160-180 Hz. We
added a Lorentzian component in order to try to improve the fit by taking this excess into
account and found a peak of significance between 3.8 and 4.3$\sigma$ in H2 and H2+H3 respectively. 
Significances of detections are computed from the normalization of the Lorentzian components
in power units.
The results of these new fits are reported in Table \ref{tab:2PL+L}.

%                                                       One column table
%-----------------------------------------------------------------------
\begin{table}
\begin{tabular}{lccccc}
\hline
Region & Centroid              & FWHM                          & rms  & $\chi_\nu^2$ & Significance  \\
               & (Hz)                      & (Hz)                              &   (\%) &              & (\# $\sigma$) \\
\hline
H2       &  164$^{+7}_{-12}$&  91$^{+37}_{-24}$    &  3.9$\pm$0.5     &  1.08  &   3.8      \\
H3       &   165 (FIX)              & 87.5 (FIX)                   & $<$1.1                &   1.07  &  --- \\
H2H3  &  166$\pm$7           &  84$^{+32}_{-20}$    &  3.7$\pm$0.4    &   1.09  &   4.3      \\
\hline
\end{tabular}
\caption[]{\footnotesize Best fit parameters for the Lorentzian component for the $\theta$ observations.
For H3, the quoted rms is a 3$\sigma$ upper limit.}
\label{tab:2PL+L}
\end{table}
%-----------------------------------------------------------------------

For H2+H3 we have a 4.3$\sigma$ detection; an F-test yields a probability of a chance improvement in the fit of less than $10^{-6}$.
The corresponding 166 Hz feature is broad, with a coherence factor 
$Q =\nu_c / \Delta = 2.0^{+0.5}_{-0.8}$, where $\nu_c$ is the centroid frequency and $\Delta$ is the
FWHM.
For H3, we fixed centroid and width to an average between the H2 and H2+H3 detections and
obtained a 3$\sigma$ upper limit to the rms of 1.1\% (see Table \ref{tab:2PL+L}).
Similarly, we obtain a 3$\sigma$ upper limit of 0.8\% for region H1.

%==============================================================

\subsection{The 1997 September observations}
\label{sect.remillard}

In order to compare our results with those presented by Remillard et al. (2002b),
we analyzed all the 14 RXTE/PCA observations made in the time interval 1997
September 5-29, corresponding to those selected and analyzed by Remillard et al. (2002b) (see Table
\ref{tab_Remillard} for a log of these observations and for an identification
of the class of variablity according to Belloni et al. 2000). From
Table \ref{tab_Remillard}, it is evident that this selection includes observations corresponding to different variability classes.

\begin{table}
\begin{tabular}{llc}
\hline
\textbf{Obs. ID} & \textbf{Start Date} & \textbf{Class}   \\
\hline
20402-01-45-02   &1997 Sep 5  4:50&  $\theta$        \\
20402-01-45-00   &1997 Sep 7 15:25&  $\beta$    \\
20402-01-45-01   &1997 Sep 7 22:05&  $\beta$    \\
20402-01-45-03   &1997 Sep 9 5:47& $\beta$    \\
20402-01-46-00   &1997 Sep 11 9:45& $\beta$    \\
20186-03-02-00   &1997 Sep 15 4:21&  $\theta$        \\
20186-03-02-01   &1997 Sep 15 18:48&  $\theta$        \\
20186-03-02-02   &1997 Sep 16 2:34&  $\theta$        \\
20186-03-02-03   &1997 Sep 16 8:18&  $\theta$        \\
20186-03-02-04   &1997 Sep 16 18:46&  $\theta$   \\
20186-03-02-05   &1997 Sep 17 1:09&  $\chi_4$        \\
20186-03-02-06   &1997 Sep 18 2:40& $\chi_4$        \\
20402-01-47-01   &1997 Sep 19 0:00&  $\chi_4$        \\
20402-01-48-00   &1997 Sep 29 14:01&  $\chi_4$        \\
\hline
\end{tabular}
\caption[]{\footnotesize Observations analyzed by Remillard et al. (2002b).}
\label{tab_Remillard}
\end{table}
%_______________________________________________________________________

\begin{figure}
\begin{center}
\includegraphics[angle=0,width=8.5cm]{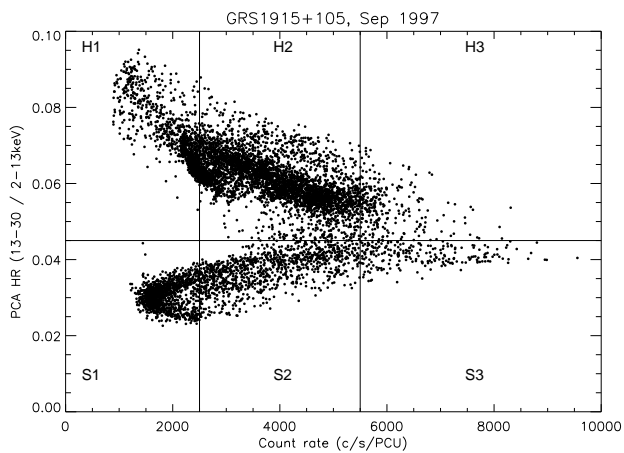}
\end{center}
\caption{Hardness-intensity diagram for GRS 1915+105 using RXTE/PCA observations
  on 10 days during 1997 September 5-29. The parameters used to produce the diagram are the same as in Fig. \ref{theta}. The plot is divided in six regions according to Remillard et al. (2002b).}
\label{remillard}
\end{figure}

In Figure \ref{remillard}, we show the total hardness-intensity diagram,
as obtained from the complete data-set. 
Notice the difference with Fig. \ref{theta}: the larger scatter of the points in this figure is caused by the mix of observations belonging to different variability classes. 
We used the same time resolution, energy
ranges and normalization to PCU2 as before. We did not subtract the background contribution to the points: this is a negligible effect on the total count rate, while resulting only in a slight upward shift of the points due to the increase of hardness.
As we did in the previous section, we divided the HID in six sections, and we averaged the power spectra for the two high-rate hard sections (H2 and H3) and for the union of the two sections (H2+H3). 

%                                                      One column figure
%-----------------------------------------------------------------------
%-----------------------------------------------------------------------

As we did for the $\theta$-class observation, we first fitted the high-frequency portions of these spectra (30-1000 Hz) with a simple power-law model (plus a constant for the Poissonian contribution).
The reduced $\chi_\nu^2$ values we obtained are 1.42, 1.07 and 1.50 for H2, H3 and H2+H3, respectively.

\begin{figure}
\begin{center}
\includegraphics[angle=0,width=8.5cm]{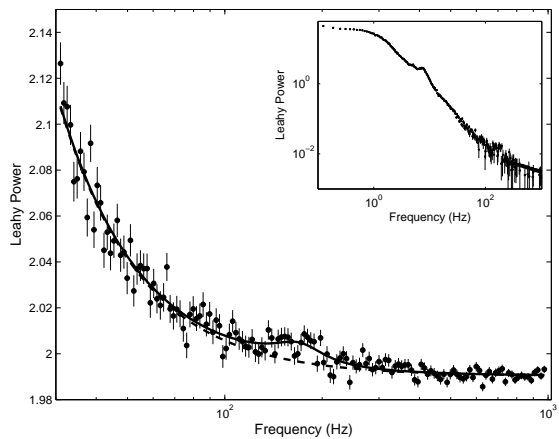}
\end{center}
\caption{Power spectrum for the H2+H3 selection for the 1997 observations. 
The main panel shows the high-frequency portion. The thick line represents the
best fit model (see text), while the dashed line is the same model with the Lorentzian component
excluded. The inset shows the full power spectrum, after subtraction of the best-fit Poissonian
level.}
\label{pds_1997}
\end{figure}

For H2 and H2+H3, where $\chi_\nu^2$ was higher than unity, we then added a Lorentzian component to account for the residuals between 100 and 200 Hz and obtained the best fits shown in Tab.\ref{tab:results}. The power spectrum corresponding to H2+H3 is shown in Fig.
\ref{pds_1997}.
For H3, we fixed centroid and width to an average between the H2 and H2+H3 detections and
obtained a 3$\sigma$ upper limit to the rms (see Table \ref{tab:results}).

%                                                       One column table
%-----------------------------------------------------------------------
\begin{table}
\begin{tabular}{lccccc}
\hline
Region & Centroid & FWHM & rms  & $\chi_\nu^2$ & Significance  \\
              & (Hz)         & (Hz)      &   (\%)&                            & (\# $\sigma$) \\
\hline
H2     & 166$\pm$8                 & 60$^{+28}_{-18}$     &  2.8$\pm$0.5   & 1.08             &       2.97    \\
H3     & 167 (FIX)                      & 70 (FIX)                       & $<$1.0             &   1.04            &   ---            \\
H2H3& 168$^{+7}_{-11}$       & 80$^{+102}_{-27}$   &  3.5$\pm$0.5   & 1.23            &       3.22    \\
\hline
\end{tabular}
\caption[]{\footnotesize Best fit parameters for the Lorentzian component for the 1997 data-set (Section \ref{sect.remillard}). For H3, the quoted rms is a 3$\sigma$ upper limit.}
\label{tab:results}
\end{table}
%-----------------------------------------------------------------------

\section{Discussion}
\label{disc}

We investigated the presence of high-frequency QPOs in a particular class of observations of
GRS 1915+105, class $\theta$ from Belloni et al. (2000), characterized by the presence of a steep
hard component in the energy spectrum during state-C intervals. State C is the one where type-C
QPOs are always observed (see Belloni et al. 2000).
We find evidence for the presence
of a broad peaked component modeled with a Lorenzian with centroid frequency $\nu_c$=166$\pm$7
FWHM  $\Delta$=84$^{+32}_{-20}$. In the 13-30 keV band, the integrated fractional rms of this
component is 3.7\%. The quality factor $Q=2.0^{+0.5}_{-0.8}$.
Comparing our results with those of Remillard et al. (2002b), we confirm the presence of a feature
around 170 Hz: however, our peak is stronger, as theirs had a fractional rms of 1.6\% in the same energy band, and our $Q$ value is substantially lower (2 vs. 5--7). Even using the same data set, we do not confirm the presence of  significant peaks at higher frequencies. The reason of these discrepancy is unclear, although it could be related to the different technique to account for the Poissonian noise level, for which we took a more conservative approach. 
The difference in width cannot be attributed to the largest sample of observations analyzed here, which 
could include more variations in frequency, as 
it is also found from the same sample used by Remillard et al. (2002). The difference in
fractional rms is probably directly related to the difference in width, since it is the integral of the 
Lorentzian function used for the fit. The difference in removal of the Poissonian contribution
could have played a role as well.

To date, most high-frequency QPOs detected in BHCs were observed during states which excluded
the presence of type-C QPOs. In this case, a clear type-C QPO is observed  (see Fig. \ref{pds_theta} and \ref{pds_1997}), as it is normal for the state C of GRS 1915+105 (see Belloni et al. 2000). Interestingly, HFQPOs are detected together with type-C QPOs also in the 2005 outburst of GRO J1655--40 (Homan,  priv. comm.). These detections and the one reported here have in common the fact that in both cases the detection corresponds to the softest spectra associated to type-C QPOs (see Fig. \ref{simone2}).
Therefore, although the mutual exclusion of type-C and HF QPOs does not strictly hold, the latter
are confirmed to be detected in correspondence of a small range in spectral parameters.

In summary, the confirmed detections of HFQPOs (with frequencies $>$20 Hz) in GRS 1915+105 to date
are the following:
\begin{itemize}
   \item {\it 27 Hz}: detected in one observation on 1996 March 14 (variability class $\gamma$) only in correspondence of the minima in the slow (period $\sim$15 s) oscillation (Belloni  et al. 2001).
   
   \item {\it 41 Hz}: detected in a set of observations in the period 1997 July-November, all of variability class $\gamma$ (Strohmayer 2001b).
   
   \item {\it 65-67 Hz}: detected in two observations from 1996 April 20 and May 5 (Morgan et al. 1997).
The other detections from this period (Morgan et al. 1997) have significances well below 3$\sigma$ (Belloni et al. 2001). Both observations belong to class variability $\gamma$. The strong 65.5 Hz QPO from May 5 shows very strong spectral changes with the $\sim$15 s slow oscillation (Belloni et al.  2001).
   
   \item {\it 69 Hz}: detected together with the 41 Hz QPO by Strohmayer (2001b) in the 1997 $\gamma$-class observations. Although their frequency is not formally compatible with 65-67 Hz, this QPO is
   most likely the same as the ones found by Morgan et al. (1997).
   
   \item {\it 166 Hz}: detected by Remillard et al. (2002b) and confirmed in this work through a more
   homogeneous and complete selection of observations. It is observed in the average of a rather large number of observations from 1996 June to 1997 October. All these observations are of variability class $\theta$, at variance with the previous ones, which were all of class $\gamma$. We do not find significant peaks at higher frequencies. As shown before, unlike the previous ones, this feature is rather broad.
   
\end{itemize}

To date there is no record of a 113 Hz oscillation as reported by McClintock \& Remillard (2005). Identifying the 65-67 Hz oscillations with the 69 Hz one, we have {\it four} different characteristic frequencies. The only ones that were ever observed at the same time were the 41 Hz and 69 Hz, while the others seem to be mutually exclusive. The values 27:41:69 are roughly in ratio 2:3:5, 
while in this sequence the 166 Hz peak would be at 12.3. 

\begin{figure}
\begin{center}
\includegraphics[angle=0,width=8.5cm]{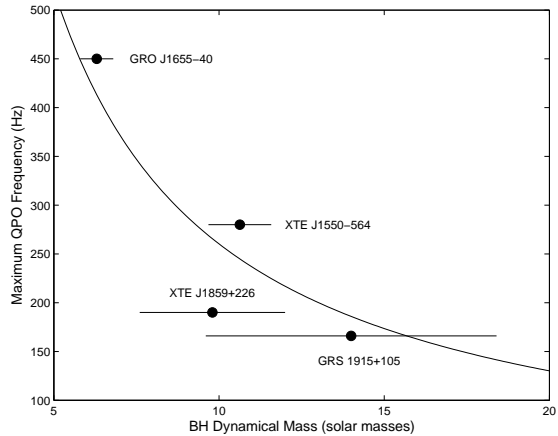}
\end{center}
\caption{Plot of the highest QPO frequency observed vs. dynamical black-hole mass for black-hole systems for which both HFQPOs and a mass estimate are available. The line is the best fit with a 1/M law.
}
\label{masses}
\end{figure}

In order to compare the frequencies observed in the different systems with the dynamical masses for the black hole one needs to follow a rigid criterion for the choice of the peaks. In absence of other indications, the obvious choice would be to consider the highest observed frequency for each system. We can therefore plot these frequencies as a function of the black-hole mass for GRO J1655-40, XTE J1550-546, GRS 1915+105 and XTE J1859+226 (masses from McClintock \& Remillard 2005; the other two systems with high-frequency QPOs, H 1743-322 and XTE J1650-500, do not have a dynamical estimate for the mass of the black hole). 
The resulting plot can be seen in Fig. \ref{masses}, where we also plot the best-fit 1/M relation. Although the fits is by no means good, the general trend is the expected one.

The QPO reported here is observed in the C state of GRS 1915+105, which corresponds to the 
Hard-Intermediate State of other black-hole transients (see Fender \& Belloni 2004; Belloni et al. 
2005; Homan \& Belloni 2005).
Most models for the production of these oscillations associate the highest-frequency peak to the keplerian time scale at the innermost stable orbit around the black hole. 
Recently, multi-wavelength observations of GX 339-4 in its hard state indicated that the thermal
accretion disc was not truncated (Miller et al. 2006). From Fig. \ref{simone2} we can see that
region H2 corresponds to an inner disc radius already at its minimum value, when 
a bright disc component is observed at all times, although it is not the dominating one.
At any rate, HFQPOs
have a hard spectrum, being observed at high energies (see Morgan et al. 1997) where the contribution of the  thermal disc component is negligible. Therefore, no direct association should be made with the thermal component. 
Interestingly, this is the highest frequency observed in GRS 1915+105 and it corresponds
to a harder energy spectrum  than that associated to the other detections. This 
seems to exclude an inverse relation between hardness and  QPO frequency.

Finally, we note that all detections of narrow ($Q \geq$2) high-frequency features from black-hole candidates are observed when a thermal disc component is present in the energy spectrum, together
with a rather steep hard component. No such detection corresponds to the high/soft state, where the hard component is very weak, nor to the low/hard state, where the hard component is flat and the disc component is weak or absent in the PCA energy band. All detections can be associated to intermediate
states (see Homan \& Belloni 2005). However, these features are so weak and elusive that from the
observations it is not clear what precise conditions are associated to their presence.

%\section*{Acknowledgments}

\label{lastpage}


\begin{thebibliography}{99}

\bibitem{AbrKlu01}
Abramowicz, M., Kluzniak, W., 2001, A\&A, 2001, L19

\bibitem{}
Belloni, T., Hasinger, G., 1990, A\&A, 230, 103

Belloni, T., M\'endez, M., S\'anchez-Fern\'andez, C., 2001, A\&A, 372, 551

\bibitem[\protect\citeauthoryear{Belloni et al.}{2000}]{b00}
Belloni, T., Klein-Wolt, M., M\'endez, M., van der Klis, M., van Paradijs, J., 2000, A\&A, 355, 271

\bibitem{}
Belloni, T., Homan, J., Casella, P., van der Klis, M., Nespoli, E., Lewin, W.H.G., 
Miller, J.M., M\'endez, M., 2005, A\&A, 440, 207

\bibitem{}
Casella, P., Belloni, T., Stella, L., 2005, ApJ, 629, 403

\bibitem{}{}
Castro-Tirado, A.J., Brandt, S., Lund, N., 1992, IAUC, 5590

\bibitem[Cui et al. 2000]{Cuietal00}
Cui, W., Shrader, C. R., Haswell, C. A., \& Hynes, R. I., 2000, ApJ, 535, L123

\bibitem{}
Fender, R.P., Belloni, T., 2004, ARA\&A, 42, 317

\bibitem{}
Homan, J., Belloni, T. 2005, Ap\&SS, 300, 107

\bibitem{}
Homan, J., Wijnands, R., van der Klis, M., Belloni, T., van Paradijs, J., 
Klein-Wolt, M., Fender, R.P., M\'endez, M., 2001, ApJSuppl., 132, 377

\bibitem[Homan et al. 2003]{Homanetal03}
Homan, J., Klein-Wolt, M., Rossi, S., Miller, J. M., Wijnands, R., Belloni,
T., van der Klis, M., Lewin, W. H. G., 2003, ApJ, 586, 1262

\bibitem{}
Homan, J., Miller, J.M., Wijnands, R., van der Klis, M., Belloni, T., 
Steeghs, D., Lewin, W.H.G., 2005, ApJ, 623, 383

\bibitem{Klein2004}
Klein-Wolt, M., Homan, J., van der Klis, M., 2004, Nucl. Phys. Proc. Suppl., 132, 381

\bibitem{}
Markwardt, C.B., Swank, J.H., Taam, R.E, 1999, ApJ, 513, L37

\bibitem{}
McClintock, J.E., Remillard, R.A., 2005, in ``Compact stellar X-ray sources'', W.H.G. Lewin \&
M. van der Klis Eds., Cambridge Univ. Press, Cambridge, in press (astro-ph/0306213)

\bibitem{}
Migliari, S., Belloni, T., 2003, A\&A, 404, 283

\bibitem{}
Miller, J.M., Wijnands, R., Homan, J., Belloni, T., Pooley, D., Corbel, S., 
Kouveliotou, C., van der Klis, M., Lewin, W.H.G., 2001, ApJ, 563, 928

\bibitem{}
Miller, J.M., Homan, J., Steeghs, D., Rupen, M., Hunstead, R.W, Wijnands, R., 
Charles, P.A., Fabian, A.C., 2006, MNRAS, submitted (astro-ph/0602633)

\bibitem{}
Mirabel, I.F., Dhawan, V., Chaty, S., Rodr\'\i guez, L.F., Mart\'\i, J., Robinson, C.R., 
Swank, J.H., Geballe, T.R., 1998, A\&A, 330, L9

\bibitem{}
Mitsuda, K., Inoue, H., Koyama, K., Makishima, K., Matsuoka, M., 
Ogawara, Y., Suzuki, K., Tanaka, Y., Shibazaki, N., Hirano, T., 
1984, PASJ, 36, 741

\bibitem{}
Morgan, E.H., Remillard, R.A., Greiner, J., 1997, ApJ, 482, 993

\bibitem{}
Remillard, R.A., McClintock, J.E., Sobczak, G.J., Bailyn, C.D., Orosz, J.A., 
Morgan, E.H., Levine, A.M., 1999, ApJ, 517, L127

\bibitem{}
Remillard, R.A., Sobczak, G.J., Muno, M.P., McClintock, J.E., 2002a, ApJ, 564, 962

\bibitem{}
Remillard, R.A., Muno, M., McClintock, J., Orosz, J., 2002b, in ``New Views on Microquasars", Eds. Ph.
Durouchoux, Y. Fuchs, J. Rodriguez, Center for Space Physics, Kolkata, India, p.49

\bibitem[Strohmayer 2001a]{Strohmayer01a}
Strohmayer, T., 2001a, ApJ, 552, L49

\bibitem[Strohmayer 2001b]{Strohmayer01b}
Strohmayer, T., 2001b, ApJ, 554, L169

\bibitem[van der Klis 2005]{vanderKlis05}
van der Klis, M.,  2005, to appear in "Compact Stellar
X-ray Sources", eds. W.H.G. Lewin and M. van der Klis, Cambridge University
Press, Cambridge (astro-ph/0410551)

%\bibitem[]{}

\end{thebibliography}
\end{document}